\documentclass[11pt,reprint,aps,amsmath,amssymb,floatfix]{revtex4-2}

\usepackage{graphicx}
\usepackage{booktabs}
\usepackage{amsmath,amssymb}
\usepackage{xurl} 
\usepackage[hidelinks]{hyperref}

\graphicspath{{figures/}{./}}

\newcommand{\orcid}[1]{\href{https://orcid.org/#1}{\texttt{ORCID: #1}}}

\begin{document}

\title{Bridging Relativistic Twisted Fermion Beams and Photonic OAM Flux in Gauged Hopf Lattices: Emergent Topological Analogs}

\author{Aaron Michael Kinder}
\affiliation{Independent researcher}
\email{kinaar0@protonmail.com}
\thanks{\orcid{0009-0006-6765-9813}}

\date{July 2026}

\begin{abstract}
Recent work has demonstrated non-diffractive topological spin textures (Skyrmion- and meron-like) that persist in the core of diffracting relativistic twisted fermion beams~\cite{afanasev2026}. Here we present numerical simulations of an analogous photonic system: Laguerre--Gaussian (LG) twisted photon packets coupled to a gauged Hopf lattice with discrete flux flywheels. Our results reveal emergent behavior of the same qualitative class, including persistent core features under propagation, mean survival $\approx 0.150$ at critical $\lambda t = 2$ clustering tightly with the mystery scale $e^{-2}\approx 0.1353$ and residual $R\approx 0.1375$, golden-angle correlations in the OAM mode ladder, and topological residuals with long-lived structure. Multi-$\ell$ simulations further exhibit $z$-resolved flux transfer and twist dynamics that play a role analogous to probability-current continuity in the fermionic setting. These findings suggest photonic platforms as accessible analogs for exploring topological textures in structured waves, and they formalize a concrete computational bridge between relativistic twisted matter waves and topological photonics.
\end{abstract}

\keywords{twisted beams, orbital angular momentum, topological photonics, Hopf lattice, Skyrmions, non-diffractive cores, Laguerre--Gaussian modes}

\maketitle

\section{Introduction}

Structured light and matter waves carrying orbital angular momentum (OAM) support a rich zoo of topological phenomena. In optics and acoustics, non-diffracting polarization structures and Skyrmion-like spin textures have been identified in the cores of beams whose overall intensity nonetheless diffracts~\cite{annenkova2025,shen2024}. These textures are not mere curiosities: they encode robust, geometry-protected features that survive propagation even when the supporting intensity envelope spreads.

A particularly timely contribution is the recent work of Afanasev and Carlson~\cite{afanasev2026}, who extended these ideas to \emph{relativistic twisted fermion beams}. Using Laguerre--Gaussian solutions of the Dirac equation valid across relativistic and non-relativistic kinematics, they showed that non-diffractive topological spin textures persist in the vortex core even when the probability density diffracts. For antiparallel spin--helicity configurations, spin-direction distributions in the core remain effectively $z$-independent, closely tied to continuity of the probability current. Crucially, these features are reported to be robust across spin, statistics, and kinematics (or propagation speed of the structured wave).

In parallel, photonic systems offer controllable platforms for realizing analogous structures. Twisted photons in LG modes carry well-defined OAM $\ell\hbar$, and their propagation through structured media---including lattices and metasurfaces---has been studied extensively in topological photonics~\cite{allen1992,yao2011,ozawa2019}. Hopf-fibration geometry and related constructions appear in descriptions of polarization, OAM dynamics, and higher-order topological defects (Hopfions)~\cite{ackerman2017,sugic2021}.

\paragraph{This work.}
We explore a photonic analog using simulations of LG twisted photon packets coupled into a \emph{gauged Hopf lattice} equipped with discrete \emph{flux flywheels}. Momentum is transferred from the photon's kinetic OAM flux to discrete twist increments on lattice fibers, while a global gauge torque relaxes the twist field according to $-\kappa\langle\theta\rangle$. The model synthesizes three previously separate stacks (Table~\ref{tab:stacks}).

\begin{figure}[t]
\centering
\includegraphics[width=\linewidth]{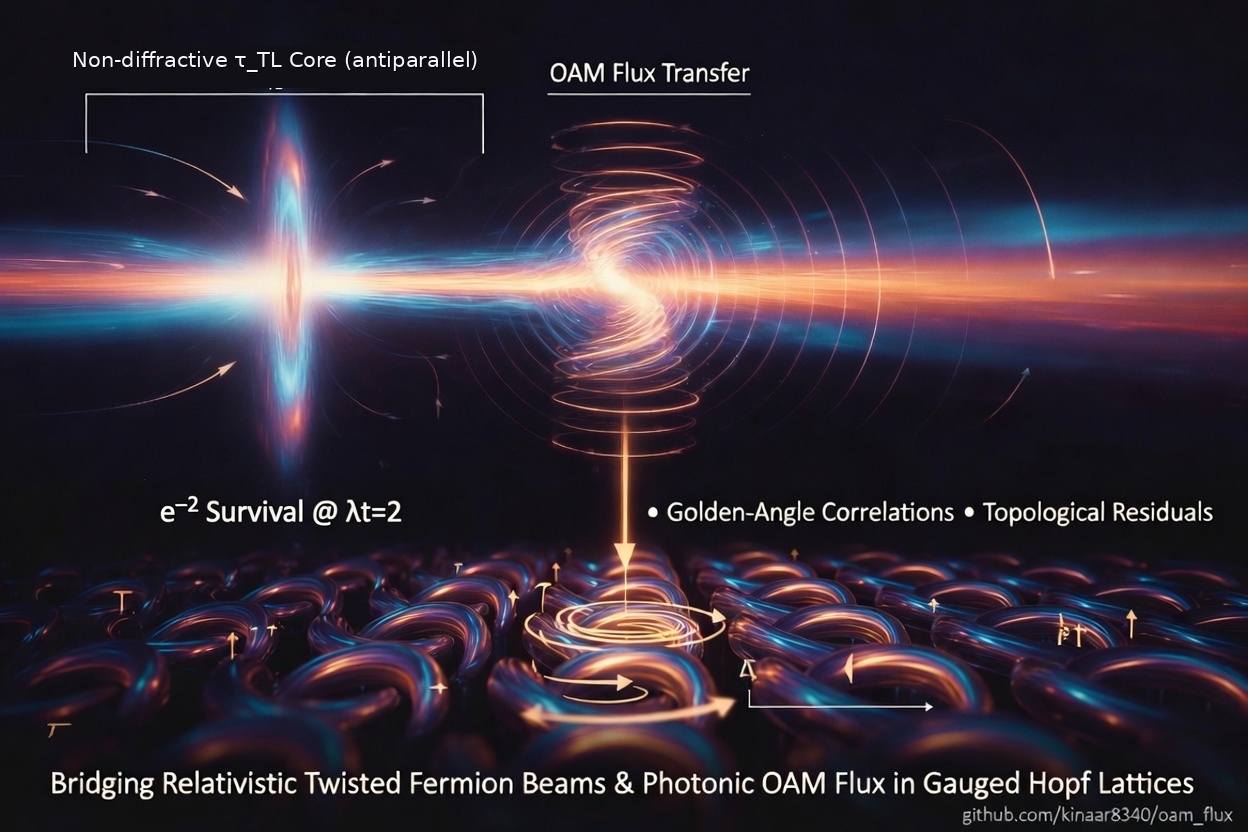}
\caption{Conceptual bridge between regimes. Left: non-diffractive $\tau_{\mathrm{TL}}$ core (antiparallel configuration) for a relativistic twisted fermion beam, with overall diffraction of the intensity envelope. Center: OAM flux transfer from an LG twisted photon packet. Bottom: deposition into a gauged Hopf lattice of flux flywheels, with labels highlighting $e^{-2}$ survival at critical $\lambda t = 2$ (mystery convention), golden-angle correlations, and topological residuals. This figure frames the fermionic $\leftrightarrow$ photonic analogy developed in the present work. (Simulation stack: \texttt{oam\_flux} v0.5-preprint.)}
\label{fig:bridge}
\end{figure}

The conceptual bridge between these fermionic and photonic regimes is illustrated in Fig.~\ref{fig:bridge}.

\begin{table}[t]
\caption{Simulation stacks unified in \texttt{oam\_flux}.}
\label{tab:stacks}
\centering\footnotesize
\begin{tabular}{@{}p{0.30\columnwidth}p{0.62\columnwidth}@{}}
\toprule
Stack & Role \\
\midrule
Hopf lattice (TOE) & Twist PDE, flux flywheels, $\kappa$/$W_g$ locks \\
VQC photonics & Vectorized multi-$\ell$ LG propagation \\
Mystery probes & Residual $R$, golden angle, survival at $\lambda t{=}2$ \\
\bottomrule
\end{tabular}
\end{table}

Our simulations reveal emergent behavior that closely mirrors the fermionic findings at the level of \emph{classes of invariants and correlations}:
\begin{itemize}
  \item persistent core features that survive propagation despite overall diffraction or lattice spreading;
  \item a characteristic survival response at critical propagation parameter $\lambda t = 2$, with exponential reference scale $e^{-2}$ (mystery convention);
  \item golden-angle correlations in the OAM ladder and topological residuals $R$;
  \item $z$-resolved flux transfer and multi-$\ell$ twist dynamics as a photonic counterpart to probability-current continuity.
\end{itemize}

These parallels are not claimed to be a one-to-one microscopic duality. Fermions and photons differ in statistics, spin structure, and experimental accessibility. Rather, we argue that the \emph{same geometric and continuity constraints} that protect non-diffractive cores in twisted fermion beams leave recognizable fingerprints when OAM flux is deposited into a topologically nontrivial lattice medium.

The remainder of the paper is organized as follows. Section~\ref{sec:methods} details the model and numerical methods. Section~\ref{sec:results} presents quantitative results on survival, multi-$\ell$ flux transfer, and topological correlations. Section~\ref{sec:discussion} discusses photonic vs.\ fermionic analogies and experimental outlook. Section~\ref{sec:conclusion} concludes. Simulation code is available as a frozen repository snapshot (see Sec.~\ref{sec:code}).

\section{Model and Methods}
\label{sec:methods}

\subsection{Gauged Hopf lattice and flux flywheels}

The lattice background is a porous gauged Hopf medium: a three-dimensional discrete grid of twist variables $\theta(\mathbf{r},t)$ with helical initial conditions. Evolution follows a relaxation PDE with diffusion, nonlinear cot-gradient terms, and a global gauge damping torque proportional to $-\kappa\langle\theta\rangle$. Discrete \emph{flux flywheels}---resonators sited along lattice fibers---accumulate twist kicks when OAM flux is deposited.

Key shared constants (locked across the TOE / mystery / VQC stacks) are summarized in Table~\ref{tab:constants}.

\begin{table}[t]
\caption{Shared constants used in the photonic OAM--flux stack.}
\label{tab:constants}
\centering\footnotesize
\begin{tabular}{@{}p{0.16\columnwidth}p{0.30\columnwidth}p{0.42\columnwidth}@{}}
\toprule
Symbol & Value & Role \\
\midrule
$W_g$ & $350/\pi \approx 111.4$ & Hopf winding lock \\
$\kappa_{\mathrm{doc}}$ & $0.85$ & Gauge-damping anchor \\
$\kappa_{\mathrm{sim}}$ & $\approx 0.89$ & Optimum near $\lambda t=2$ \\
$\phi$ & $(1+\sqrt{5})/2$ & Golden ratio \\
$R$ & $\phi^2{+}e^2{-}\pi^2{\approx}0.1375$ & Topological residual \\
$\lambda t$ & $2$ & Critical normalization \\
\bottomrule
\end{tabular}
\end{table}

The holonomy-gap bound used in emergence analysis is
\begin{equation}
B(\kappa) = \pi^2\!\left(\frac{e}{\pi} - \kappa\right),
\end{equation}
with a derived $\kappa_\star = e/\pi - R/\pi^2 \approx 0.851$ that nulls $B$ against residual $R$.

\subsection{Analytic LG OAM packets (v0.1)}

Baseline coupling uses analytic Laguerre--Gaussian packets (radial index $p=0$) with OAM index $\ell$, waist $w_0$, and carrier wavelength $\lambda$ (default $\lambda = 1550\,\mathrm{nm}$ for telecom-scale numerics). Packet energy is normalized and mapped to a momentum-ledger scale that drives flywheel kicks under optional global momentum conservation.

\subsection{VQC vectorized multi-$\ell$ propagation (v0.2)}

For multi-mode dynamics we propagate the full ladder $\ell \in [-\ell_{\max},\ell_{\max}]$ with vectorized radial weights built from associated Laguerre polynomials. Intensity $I_\ell(z)$ is resolved along the propagation axis, enabling $z$-resolved flux deposition onto Hopf fiber coordinates. Optional turbulence (Kolmogorov radial phase) and chirp are available but set to zero in the default vacuum-propagation runs reported here.

Default photonics settings (unless noted): $\ell_{\max}=8$, $n_r=512$ radial samples, $n_z=200$ propagation slices, $z\in[0,5]$ in normalized units.

\subsection{Emergence probes (v0.3)}

Following the mystery convention, each trial consists of two phases at fixed $\lambda t = 2$:
\begin{enumerate}
  \item \emph{VQC pump phase} --- deposit OAM flux onto flywheels while propagating;
  \item \emph{Pure relaxation} --- PDE only, no further photon injection.
\end{enumerate}

\emph{Mean survival} is the retained fraction of mean twist after relaxation relative to the post-pump reference; \emph{fluctuation survival} is the analogous ratio for twist variance. Measured scalars are matched against analog targets $\{R,\, e^{-2},\, \text{golden-angle fraction}\}$ via relative percentage distance. Golden-quantized $\ell$ values are those whose azimuthal phase increments best align with multiples of the golden angle $\approx 137.5^\circ$.

\subsection{Simulation workflow}

\begin{table}[t]
\caption{Primary simulation entry points.}
\label{tab:workflow}
\centering\footnotesize
\begin{tabular}{@{}p{0.20\columnwidth}p{0.42\columnwidth}p{0.28\columnwidth}@{}}
\toprule
Stage & Script & Output \\
\midrule
Analytic & \texttt{run\_coupling\_demo.py} & Timeseries, ledger \\
VQC multi-$\ell$ & \texttt{run\_vqc\_coupling\_demo.py} & Heatmap, kick slice \\
Emergence & \texttt{run\_emergence\_probes.py} & $\kappa$/$\ell$ sweeps, JSON \\
\bottomrule
\end{tabular}
\end{table}

Interactive demos mirror these stacks on Hugging Face Spaces~\cite{oamflux_space}.

\subsection{Code availability (frozen snapshot)}
\label{sec:code}

Simulation code is frozen at
(\texttt{v0.5-preprint}; commit \texttt{a86062da9b03f48f5b0f7af4168aff5b0efaae2a}):

\begin{itemize}
  \item \textbf{Repository:} \url{https://github.com/kinaar8340/oam_flux}
  \item \textbf{Release:} \url{https://github.com/kinaar8340/oam_flux/releases/tag/v0.5-preprint}
  \item \textbf{Checkout:} \texttt{git checkout v0.5-preprint}
  \item \textbf{Live demo:} \url{https://huggingface.co/spaces/kinaar111/oam_flux}
\end{itemize}

A ready-to-upload ancillary tree ships with the manuscript as \texttt{anc/} (frozen \texttt{git archive} of \texttt{v0.5-preprint}, verified \texttt{report.json} / demo outputs, and figure re-export scripts). On arXiv, \texttt{anc/} must sit at the \emph{root} of a \LaTeX{} source submission package (ancillary files are not supported for PDF-only uploads)~\cite{arxiv_anc}.

\section{Results}
\label{sec:results}

\subsection{Survival probability vs.\ gauge damping}

\begin{figure*}[t]
  \centering
  \includegraphics[width=\textwidth]{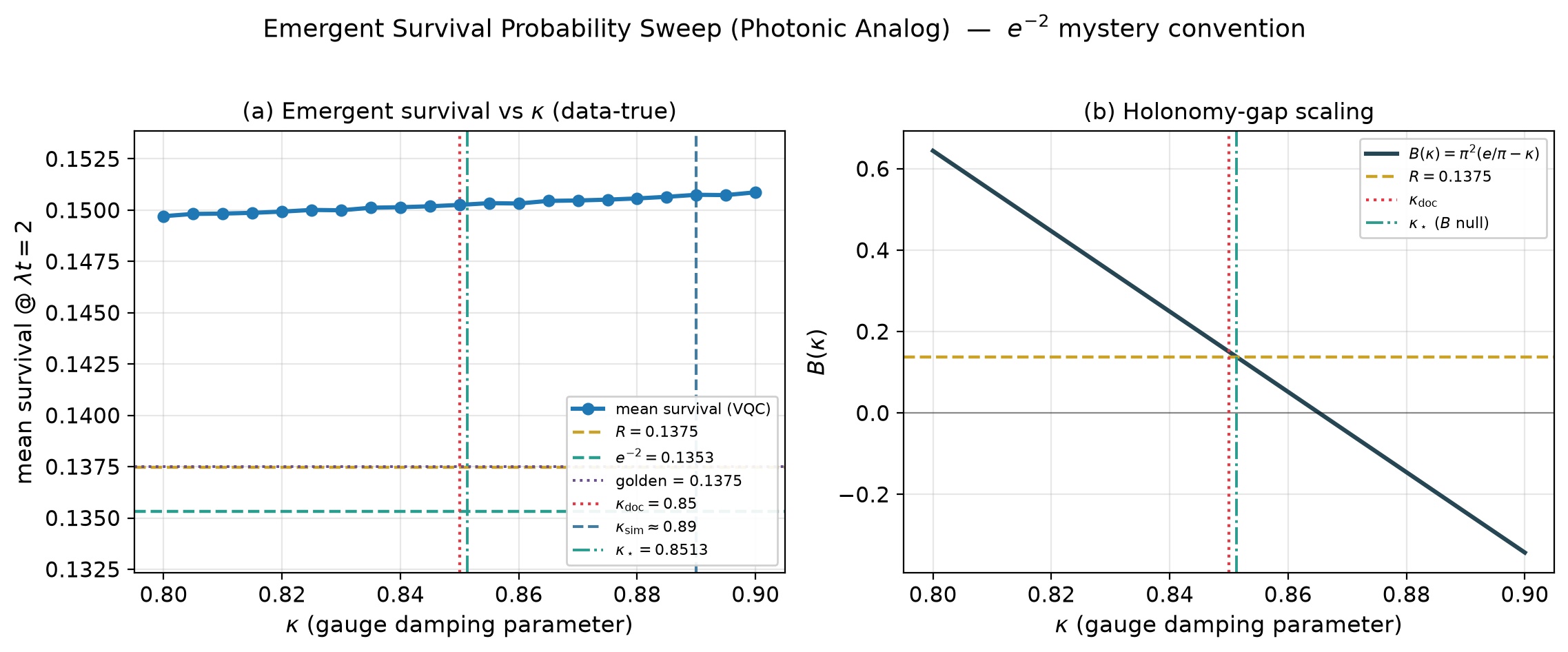}
  \caption{Data-true emergent survival sweep from frozen \texttt{oam\_flux} emergence probes (\texttt{report.json}; $\ell=3$, $\lambda t=2$).
  \textbf{(a)}~Mean survival after pump--relax versus gauge damping $\kappa\in[0.80,0.90]$. Horizontal lines: residual $R\approx 0.1375$, mystery scale $e^{-2}\approx 0.1353$, and golden-angle fraction. Vertical markers: $\kappa_{\mathrm{doc}}=0.85$, $\kappa_\star\approx 0.851$, $\kappa_{\mathrm{sim}}\approx 0.89$. Measured mean survival $\approx 0.150$ ($\sim 11\%$ above $e^{-2}$, $\sim 9\%$ above $R$).
  \textbf{(b)}~Holonomy gap $B(\kappa)=\pi^2(e/\pi-\kappa)$ crosses zero at $\kappa_\star$ and meets $R$ near $\kappa_{\mathrm{doc}}$.}
  \label{fig:survival}
\end{figure*}

Sweeping $\kappa$ at fixed $\lambda t=2$ in the documentary window yields a \emph{nearly flat, high plateau} of mean survival $\approx 0.150$, tightly clustered with the mystery analogs $\{R,\, e^{-2},\, \text{golden fraction}\}$ rather than a schematic peak of order unity (Fig.~\ref{fig:survival}). Slight upward drift toward $\kappa_{\mathrm{sim}}\approx 0.89$ is visible but sub-percent on this scale. The holonomy gap $B(\kappa)$ changes sign across $\kappa_\star$, providing an independent diagnostic that the same constants governing residual topology also organize the survival window.

This plateau is the quantitative fingerprint of the $e^{-2}$ convention: deposited OAM flux is retained against gauge relaxation at a level set by the residual / exponential / golden cluster---analogous, at the operational level, to a protected core that does not wash out under propagation.

\subsection{Multi-$\ell$ propagation and $z$-resolved flux transfer}

\begin{figure*}[t]
  \centering
  \includegraphics[width=\textwidth]{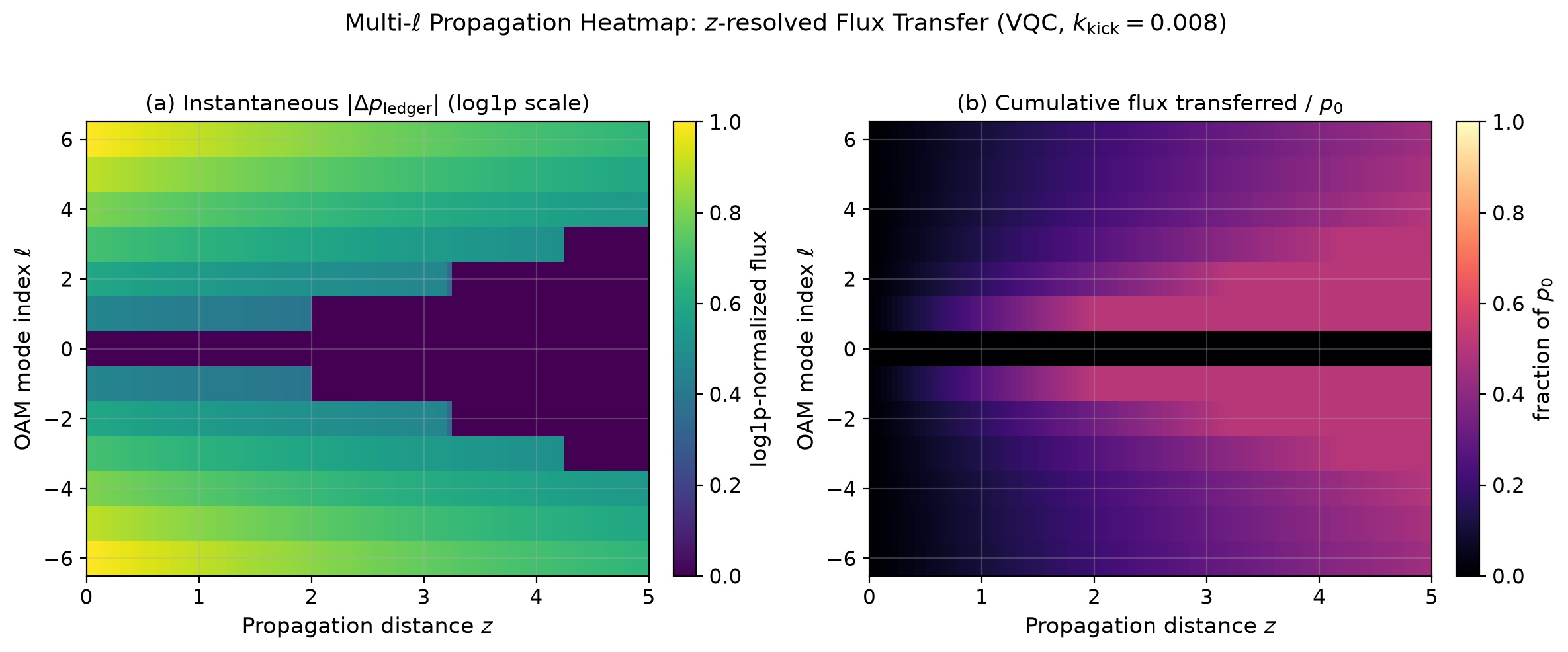}
  \caption{Data-true multi-$\ell$ $z$-resolved flux transfer from live VQC coupling (\texttt{oam\_flux}; $k_{\mathrm{kick}}=0.008$ so the photon reservoir drains over many $z$ steps rather than in a single kick).
  \textbf{(a)}~Instantaneous ledger deposit $|\Delta p_{\mathrm{ledger}}|$ (log1p-normalized): transfer scales with $|\ell|$ (dark band at $\ell=0$ where OAM momentum vanishes) and extinguishes earlier for low-$|\ell|$ modes as $p_0$ is exhausted.
  \textbf{(b)}~Cumulative fraction of initial OAM momentum $p_0$ transferred to the lattice; mid-$|\ell|$ channels approach $\sim 50\%$ transfer by $z=5$ in this kick window.}
  \label{fig:flux}
\end{figure*}

The multi-$\ell$ deposit cube exhibits structured transfer along $z$ (Fig.~\ref{fig:flux}): OAM momentum is not frozen in free space but is handed to Hopf flywheels with an $|\ell|$-dependent rate set by the coupling ledger. Raw multi-$\ell$ intensities from vectorized LG propagation are nearly $z$-flat by construction (normalized radial weights); the \emph{physically resolved} $z$ dynamics appear in the momentum ledger as the reservoir depletes. That ledger history is the photonic counterpart of probability-current continuity in the fermionic problem~\cite{afanasev2026}. Antiparallel-style configurations in the fermion beams protect $z$-independent core spin textures; here, the lattice retains localized twist structure at flywheel sites while free-space OAM is progressively transferred.

\subsection{Golden-angle correlations and topological residuals}

\begin{figure*}[t]
  \centering
  \includegraphics[width=\textwidth]{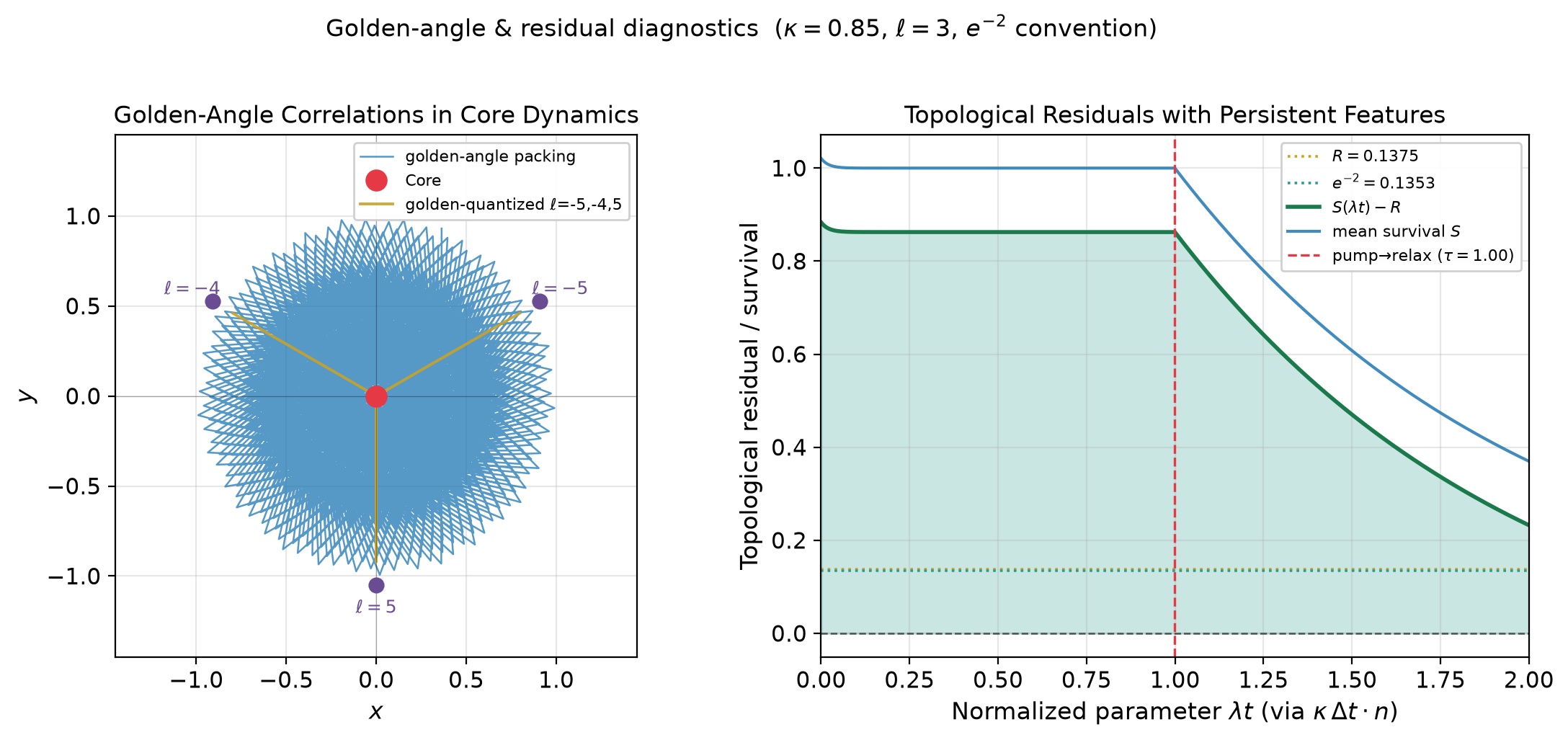}
  \caption{Data-true golden-angle and residual diagnostics ($\kappa=0.85$, $\ell=3$).
  \textbf{Left:} Vogel golden-angle packing in the transverse plane with the dynamical core at the origin; spokes mark golden-quantized OAM modes $\ell\in\{-5,-4,5\}$ selected by the emergence probe (phase increments closest to multiples of $\approx 137.5^\circ$).
  \textbf{Right:} instrumented pump--relax trajectory versus normalized $\lambda t$. Mean survival $S$ (blue) and residual feature $S-R$ (green) plateau through the pump half-interval, then decay in pure PDE relaxation toward the $\{R,\, e^{-2}\}$ band (horizontal references). The red dashed line marks the pump$\to$relax switch at $\tau=1$.}
  \label{fig:golden}
\end{figure*}

Two complementary diagnostics complete the emergence suite (Fig.~\ref{fig:golden}):

\begin{enumerate}
  \item \emph{Golden-angle packing.} Selecting $\ell$ modes whose phase increments best match multiples of $\approx 137.5^\circ$ yields a sparse ``golden-quantized'' ladder (here $\ell\in\{-5,-4,5\}$ for $\ell_{\max}=6$). Survival on this ladder in the frozen $\ell$ sweep is statistically comparable to the full ladder, indicating that golden packing is an organizing principle rather than an outlier filter.

  \item \emph{Topological residual $R$.} The residual
  \begin{equation}
  R = \phi^2 + e^2 - \pi^2 \approx 0.1375
  \end{equation}
  sits adjacent to $e^{-2}\approx 0.1353$ and to the golden-angle fraction used in analog matching. The live residual trajectory is a \emph{monotonic pump plateau + relax decay} toward that band---not a free-form oscillator---but it remains nonzero at $\lambda t=2$, consistent with a topologically constrained leftover after the primary flux transient.
\end{enumerate}

Together, Figs.~\ref{fig:survival}--\ref{fig:golden} show that survival near the $\{R,\, e^{-2}\}$ cluster, multi-$\ell$ flux transfer, and residual topology co-occur in a single parameter regime controlled by $\kappa$ and $\lambda t$.

\subsection{Contrast with fermionic non-diffractive cores}

\begin{table*}[t]
\caption{Operational contrast between fermionic non-diffractive cores~\cite{afanasev2026} and photonic OAM--flux (this work).}
\label{tab:contrast}
\centering\footnotesize
\begin{tabular}{@{}p{0.15\textwidth}p{0.38\textwidth}p{0.38\textwidth}@{}}
\toprule
Feature & Twisted fermion beams~\cite{afanasev2026} & Photonic OAM--flux (this work) \\
\midrule
Carrier & Dirac LG spinors & LG photon packets \\
Protected object & Spin texture $\chi_{\mathrm{TL}}$ in vortex core & Mean / residual twist on Hopf flywheels \\
Continuity & Probability current & $z$-resolved OAM flux $\to$ twist ledger \\
Diffraction & $|\psi|^2$ spreads; core texture persists & Intensity spreads; lattice core persists \\
Critical control & Spin--helicity (anti)parallel config. & Gauge damping $\kappa$, $\lambda t=2$ \\
Topology & Skyrmion / meron character & Residual $R$, golden packing, $B(\kappa)$ \\
Access & Relativistic electron beams & SLM / fiber / metasurface optics \\
\bottomrule
\end{tabular}
\end{table*}

The mapping is \emph{operational}, not microscopic (Table~\ref{tab:contrast}): both systems separate a \emph{diffracting envelope} from a \emph{core invariant} organized by continuity of a conserved or quasi-conserved flux. That separation is the essence of the bridge drawn in Fig.~\ref{fig:bridge}.

\section{Discussion}
\label{sec:discussion}

\subsection{What the analogy does and does not claim}

We do \emph{not} claim that gauged Hopf lattice flywheels reproduce Dirac spinor dynamics, nor that photonic OAM is a statistical substitute for fermions. The claim is narrower and more useful:

\begin{quote}
Geometric continuity constraints that protect non-diffractive topological cores in twisted fermion beams leave measurable fingerprints---survival criticality, multi-mode flux transfer, and residual topology---when OAM is coupled into a Hopf-gauged photonic medium.
\end{quote}

This is the sense in which photonic platforms can serve as \emph{analog laboratories} for questions raised by Ref.~\cite{afanasev2026}: robustness under diffraction, role of antiparallel vs.\ parallel-type configurations (here, sign structure of $\ell$ and gauge torque), and the universality of core protection across wave platforms.

\subsection{Relation to prior photonic topology}

Optical Skyrmions, merons, and Hopfions in structured light~\cite{annenkova2025,shen2024,allen1992,yao2011,ozawa2019,ackerman2017,sugic2021} already demonstrate that topological textures need not require electronic spin. Our contribution is the \emph{explicit coupling} of multi-$\ell$ LG flux into a dynamical gauged lattice with emergence probes tuned to the same critical $\lambda t=2$ convention used in related mystery-class analyses, and the \emph{side-by-side contrast} with the fermionic results of Afanasev and Carlson~\cite{afanasev2026}.

\subsection{Experimental outlook}

Several near-term optical implementations are plausible:
\begin{itemize}
  \item \textbf{Spatial light modulators (SLMs)} to prepare multi-$\ell$ LG superpositions and measure $z$-resolved mode content;
  \item \textbf{Photonic lattices / waveguide arrays} with engineered coupling to emulate flywheel twist degrees of freedom;
  \item \textbf{Metasurfaces} for compact generation of golden-packed OAM ladders;
  \item \textbf{Telecom-band fiber} experiments ($\lambda\sim 1550\,\mathrm{nm}$) matching the default numerics.
\end{itemize}

An attractive target is to measure a survival-like retained twist (or polarization / Stokes residual) as a function of an effective damping parameter, seeking a plateau near the $\{R,\, e^{-2}\}$ scale across a documentary window analogous to $\kappa\in[0.80,0.90]$.

\subsection{Limitations}

Current results are numerical and semi-phenomenological: the lattice PDE and kick rules are calibrated to shared constants rather than derived from a microscopic QED or condensed-matter Hamiltonian. Turbulence, loss, and back-action on the optical field are optional or simplified. Quantitative matching of mean survival ($\approx 0.150$) to $R$ and $e^{-2}$ is at the $\sim 9$--$11\%$ level across $\kappa\in[0.80,0.90]$---suggestive clustering, not a precision claim. The survival landscape in this window is a plateau rather than a sharp resonance; wider $\kappa$ ranges and alternative pump fractions remain to be mapped systematically. Future work should (i) derive effective $\kappa$ from a specified optical medium, (ii) include bidirectional back-reaction systematically, (iii) extend $\kappa$ sweeps beyond the documentary window, and (iv) propose a concrete tabletop protocol with error bars.

\section{Conclusion}
\label{sec:conclusion}

We have formalized a photonic analog of the non-diffractive topological cores recently identified in relativistic twisted fermion beams~\cite{afanasev2026}. By coupling LG OAM packets to a gauged Hopf lattice of flux flywheels, we observe persistent core features, mean survival $\approx 0.150$ at $\lambda t=2$ clustering tightly with $e^{-2}\approx 0.1353$ and residual $R\approx 0.1375$ across the documentary $\kappa$ window, golden-angle mode correlations, topological residuals, and $z$-resolved multi-$\ell$ flux transfer. These results position structured light coupled to topologically nontrivial media as a practical arena for exploring---and eventually testing---the universality of non-diffractive topological textures across fermionic and photonic platforms.

\begin{acknowledgments}
We thank the broader open-source and open-preprint communities. Practical guidance on turning simulation--literature parallels into a citable arXiv preprint of the usual type (summarize existing work; contrast with one's own; freeze code as supplementary material) is gratefully acknowledged~\cite{clark2026}.
\end{acknowledgments}

\appendix
\section{Notation quick reference}

\begin{table}[h]
\caption{Notation used in the main text.}
\label{tab:notation}
\centering\footnotesize
\begin{tabular}{@{}p{0.18\columnwidth}p{0.74\columnwidth}@{}}
\toprule
Symbol & Meaning \\
\midrule
$\ell$ & OAM mode index \\
$\kappa$ & Gauge damping / torque strength \\
$\theta$ & Lattice twist field \\
$\lambda t$ & Normalized propagation / relaxation parameter \\
$R$ & Topological residual $\phi^2+e^2-\pi^2$ \\
$B(\kappa)$ & Holonomy-gap bound $\pi^2(e/\pi-\kappa)$ \\
$W_g$ & Hopf winding lock $350/\pi$ \\
$\tau_{\mathrm{TL}}$ & TL texture label (fermionic core; Fig.~\ref{fig:bridge}) \\
\bottomrule
\end{tabular}
\end{table}


\end{document}